\documentclass[prl,preprint,showpacs,groupaddress]{revtex4}

\usepackage{amsmath}
\usepackage{epsfig}
\usepackage{rotating}

\begin{document}

\title{Size effects in surface reconstructed $\langle 100 \rangle$ and 
       $\langle 110 \rangle$ silicon nanowires}

\author{R. Rurali}
\affiliation{Laboratoire Collisions, Agr\'{e}gats, R\'{e}activit\'{e},
             IRSAMC, Universit\'{e} Paul Sabatier,
             118 route de Narbonne, 31062 Toulouse cedex,
             France}
\affiliation{Departament d'Enginyeria Electr\`{o}nica,
             Universitat Aut\`{o}noma de Barcelona
             08193 Bellaterra, Spain}
\author{A. Poissier}
\author{N. Lorente}
\affiliation{Laboratoire Collisions, Agr\'{e}gats, R\'{e}activit\'{e}, 
             IRSAMC, Universit\'{e} Paul Sabatier, 
             118 route de Narbonne, 31062 Toulouse cedex, 
             France}

\date{\today}

\begin{abstract}
The geometrical and electronic structure properties of $\langle 100 
\rangle$ and $\langle 110 \rangle$ silicon nanowires in the absence
of surface passivation are studied by means of density-functional 
calculations. As we have shown in a recent publication [R.~Rurali 
and N.~Lorente, Phys. Rev. Lett. {\bf 94}, 026805 (2005)] the 
reconstruction of facets can give rise to surface metallic 
states. In this work, we analyze the dependence of 
geometric and electronic structure features 
on the size of the wire and on the growth direction.
\end{abstract}


\maketitle

\section{Introduction}
\label{sec:intro}

In recent years one-dimensional quantum systems have attracted 
great interest as building blocks for future molecular electronics 
devices. Besides proposed applications, there is also a 
fundamental interest in these systems, because as the size 
is reduced down to the nanoscale 
a description 
at the quantum level is necessary to account for most of the properties 
of the material.

Carbon nanotubes are certainly the most studied one-dimensional
quantum systems. However, the dependence of their electronic
properties on their diameter and on their chirality -~and the difficulty
of controlling such properties during growth~- might make them 
an impractical choice for nanodevice engineering. 
Semiconductor nanowires seem to obviate such an inconvenient
and they are becoming the focus of a growing research interest.
Silicon nanowires~(SiNWs) are an especially appealing choice,
as they offer the best possible interface with conventional 
silicon-based microelectronics.

SiNWs have been successfully doped with both {\em n}- and {\em p}-type
impurities~\cite{lieber1,lieber2,zhou,heath} and nanoscale logic
devices have been demonstrated~\cite{lieber_logic1,lieber_logic2}.
A growing research interest has been devoted to the synthesis
of SiNW-based heterostructures -~oriented to optoelectronics 
applications~\cite{hetero1}, as well to device fabrication~\cite{hetero2,
hetero3}~- and to the use of SiNWs as chemical sensors~\cite{lieber_sensor1,
lieber_sensor2,sensor1,sensor2}.
At present, diameters around 100~nm thick can be almost routinely 
obtained and a few examples of SiNWs below 10~nm have been 
published~\cite{holmes,wu,colemann1,colemann2}, being 1.5~nm the 
thinnest wire's diameter reported so far~\cite{ma}.

Theoretically, at first the attention
was directed to the structure of the thinnest possible SiNWs. 
Hollow structures~\cite{menon} and systems ranging from atomic 
wires to silicon cluster aggregates have been proposed~\cite{li}.
However, a wealth of experimental results~\cite{wu,ma,zhang} soon
showed that realistic SiNWs present a well-defined diamond-like 
bulk core. Since then, significant efforts have been devoted to 
the modeling of hydrogen-passivated SiNWs~\cite{h-term1,h-term2,
delley,read,uzi,mei-jin}, {\em i.e.} wires where the lateral surface 
dangling bonds are terminated with hydrogens. These are extremely 
interesting systems for theoretical modeling for different reasons:
(a) the absence of surface states allows an easier approach
to the study of all the features related to quantum 
confinement~\cite{h-term1,delley,read,mei-jin}, {\em e.g.} the energy 
band-gap is well-defined and depends only on the growth direction 
and on the wire diameter; (b) it is a good approach to realistic 
experimental conditions where different kinds of polluting agents, 
{\em e.g.} hydrogen~\cite{ma}, oxygen~\cite{wang}, are present 
and can react with the surface dangling bonds.

However, studying the surface physics of these low-dimensional
systems without passivation is of great interest because of the 
subtle interplay between an unusually large surface-to-bulk ratio 
and the effect of quantum confinement. In a previous 
publication~\cite{noi_prl} we have shown that in the absence of 
passivation, the lateral surface of a $\langle 100 \rangle$ SiNW 
undergoes a strong reconstruction which can turn the wire into 
metallic. The conductive surface states form at the \{100\} facets 
of the wire, at variance with the infinite Si(100) surface which is 
known to be semiconducting. Therefore, the diameter of the SiNW must 
play a very crucial role in the determination of the surface electronic 
structure.

In this paper we analyze the diameter dependence of the surface
reconstruction and of the associated electronic structure $\langle 
100 \rangle$ SiNWs and we extend our study to wires grown along the
$\langle 110 \rangle$ direction.

\section{Computational methods}
\label{sec:methods}

We have performed density functional theory (DFT) calculations, in the
framework of the generalized gradient approximation~(GGA)~\cite{gga}.
We have used both numerical atomic orbital~\cite{siesta1,siesta2} 
and plane-wave methods~\cite{dacapo}, combined with norm-conserving 
pseudopotentials of the Troullier-Martins type~\cite{troullier:martins} 
and with ultrasoft pseudopotentials, respectively. We have used a 
double-$\zeta$ polarized basis set~\cite{basis1,basis2,basis3} and a 
plane-wave cutoff of 20~Ry for the one-electron wave function of the 
valence electrons.

All the calculations use periodic boundary conditions, thus
the supercell size may restrict the number of observed reconstructions. 
For this reason, to explore a larger set of the phase space of the 
surface reconstruction we have also used a non-orthogonal tight-binding 
model~\cite{porezag,troca} to treat system sizes that cannot efficiently 
be managed within DFT. Even in the cases when the TB surface electronic 
structure turned out to be only qualitatively accurate, the minimum energy 
geometries proved to be very close to those obtained by DFT. 
In all the calculations discussed, the atomic positions have been
relaxed until the maximum force was lower than 0.04~eV/\AA.

We have studied wires grown along the $\langle 100 \rangle$ direction
of $\sim 8$, $\sim 11$, $\sim 15$ and $\sim 27$~\AA\ and wires grown 
along the $\langle 110 \rangle$ direction of $\sim 12$, $\sim 20$, 
and $\sim 26$~\AA, being the $\sim 8$ and $\sim 12$~\AA\ the
thinnest possible wires with a bulk core of each growth orientation.
For each diameter considered we have carried out calculations with
up to four unit cells of the unreconstructed wire. The systems with
the largest number of atoms (large diameter, many unit cells) have been
only calculated with TB.
The reciprocal space has been sampled with a converged number of
{\em k}-points, ranging from 1 to 12, depending on the size of the unit 
cell in direct space. 

After converging the atomic positions, we have recalculated the 
electronic structure with a three times thicker mesh of {\em k}-points 
and an {\em electronic temperature} approaching zero, in order to 
obtain a highly accurate band diagram, especially for what concerns 
the crossings of the Fermi level. 

\section{Results and Discussion}
\label{sec:results}

\subsection{$\langle 100 \rangle$ SiNWs}
\label{sub:100}

\begin{figure}[h]
\begin{center}
\epsfxsize=14cm
\epsffile{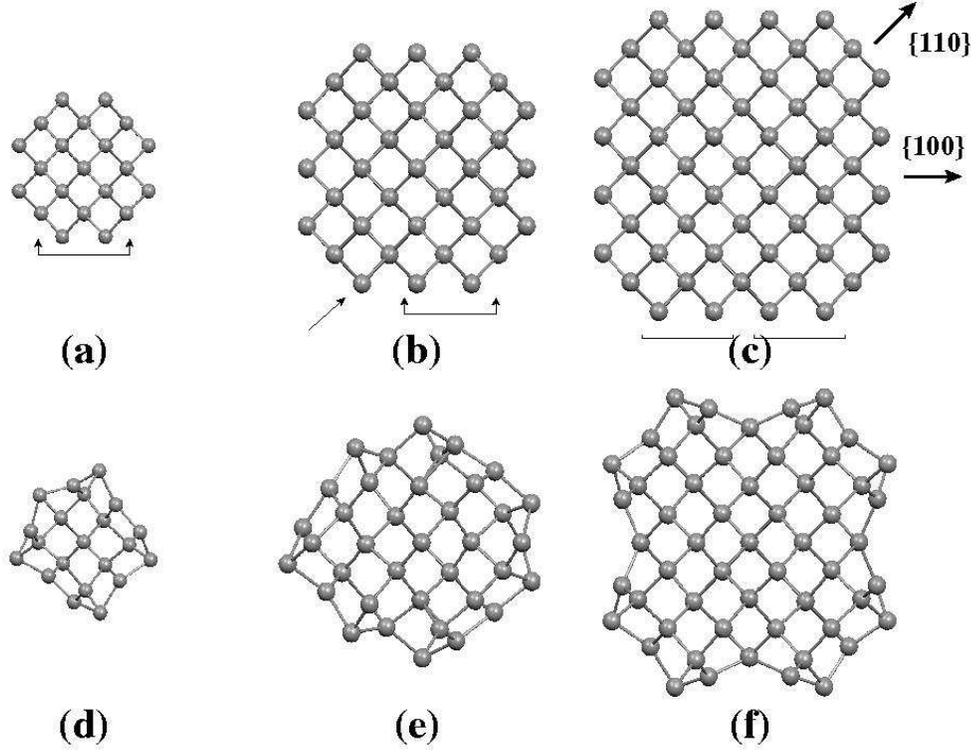}
\end{center}
\caption{(Color online) Section views of the unrelaxed [(a) to (c)] and 
         relaxed [(d) to (f)] $\langle 100 \rangle$ SiNWs with a diameter 
         of $\sim 8$\, $\sim 11$ and $\sim 15$~\AA. 
         The facet geometry is determined by the number of rows of 
         dimers (indicated by the brackets) that can form: (a,d) one-single
         row of dimers can form; (b,e) the formation of a second row of
         dimers is frustrated and the presence of an unpaired row of
         atoms results in an asymmetric relaxation of the facet; (c,f) two
         rows of dimers form, yielding a {\em trough} in the middle of
         the \{100\} facet.}
\label{fig:100_thin_sec}
\end{figure}

The first issue to deal with is the 
shape of the wire's cross-section. Although we restrict to a 
specific growth direction, different sections are possible. We have 
discussed this topic elsewhere~\cite{noi_nanotech}, concluding 
that $\langle 100 \rangle$ SiNWs favor the formation of \{100\} facets. 
On the other hand, we found that it is not clear if, at such small 
diameters, smoothing sharp angles increases or not the 
stability~\cite{noi_nanotech}. A discussion of the role of the edges 
can also be found in the work by Zhao and Yakobson~\cite{yakobson} 
or by Ismail-Beigi and Arias~\cite{arias}.
Throughout this article, we will focus our attention on wires with 
smooth angles [see Fig.~\ref{fig:100_thin_sec} (a-c)], even though 
most of the conclusions could be generalized to the case of wires 
with a square section (see Ref.~\onlinecite{noi_nanotech} for a 
complete discussion). 

\begin{figure}[h]
\begin{center}
\epsfxsize=14cm
\epsffile{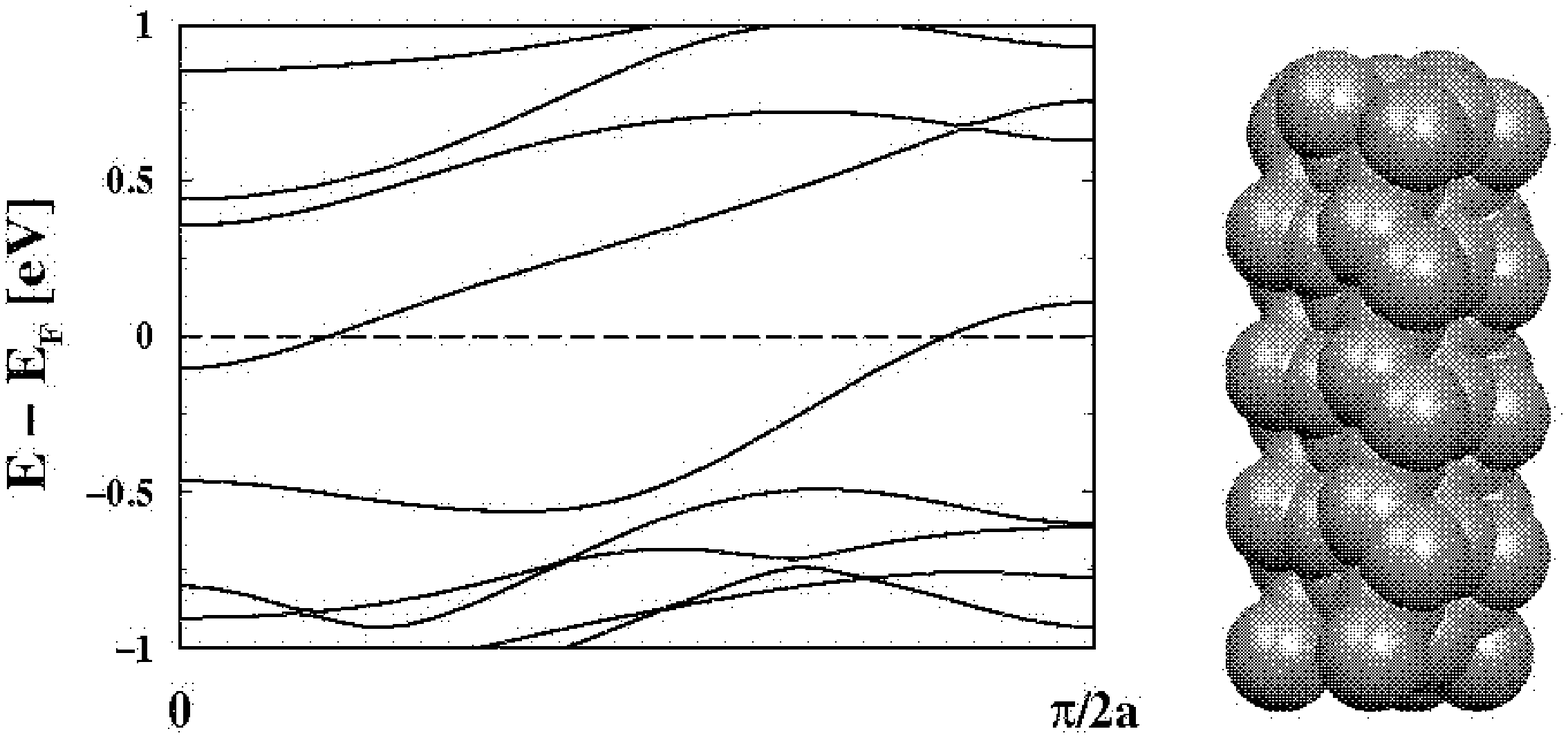}
\end{center}
\caption{(Color online) $\langle 100 \rangle$ SiNWs of $\sim8$~\AA\ 
         diameter: side view of the reconstructed (pseudo) \{100\} facet.}
\label{fig:100_8_relax_side}
\end{figure}

The first $\langle 100 \rangle$ SiNW that we have analyzed has a diameter
of $\sim 8$~\AA. This size is well below the diameter of the thinnest
SiNW grown so far ($\sim 15$~\AA, see Ref.~\onlinecite{ma}), however,
it can be a useful case study to elucidate the atomic scale mechanisms
that are capable of inducing a metallic state at the wire's surface.
We have found one only stable reconstruction which is illustrated
in Fig.~\ref{fig:100_8_relax_side}. It is characterized by 
one row of Si dimers along the {\em z}-axis. The formation of Si
dimers on the \{100\} facets is a feature common to $\langle 100 \rangle$ 
SiNWs~ and is one of the characteristic reconstruction patterns of 
the infinite Si(100) surface. 

The band structure diagram, displayed in Fig.~\ref{fig:100_8_relax_side}, 
presents two states at the Fermi energy. It is well-known 
that DFT, though giving the correct dispersion of bands, 
underestimates band-gaps. Therefore, especially in the case of 
marginal crossings of the Fermi level it is difficult to guarantee 
the metallicity of the system. An accurate band-gap estimation can 
be obtained within the GW approximation (see for instance 
Ref.~\onlinecite{mei-jin}, where the effect of quantum confinement 
in semiconducting SiNWs is studied). 
For the purpose of the present work, the gap {\em renormalization} can be modeled 
with a scissor operator which rigidly shifts upward/downward the 
states above/below a certain energy. In the present case we cannot 
exclude that the proper scissor operator would open a small gap between 
the two states that in DFT electronic structure crosses the Fermi 
level, in such a way that the 8~\AA\ SiNW would rather be a small-gap 
semiconductor~\cite{noi_prl,noi_nanotech,io_prb}. Contrary to the band-widening effect
of hydrogen-passivated wires, we notice that reconstructed wires present
a reduction of the band gap as compared to the bulk value.

\begin{figure}[h]
\begin{center}
\epsfxsize=14cm
\epsffile{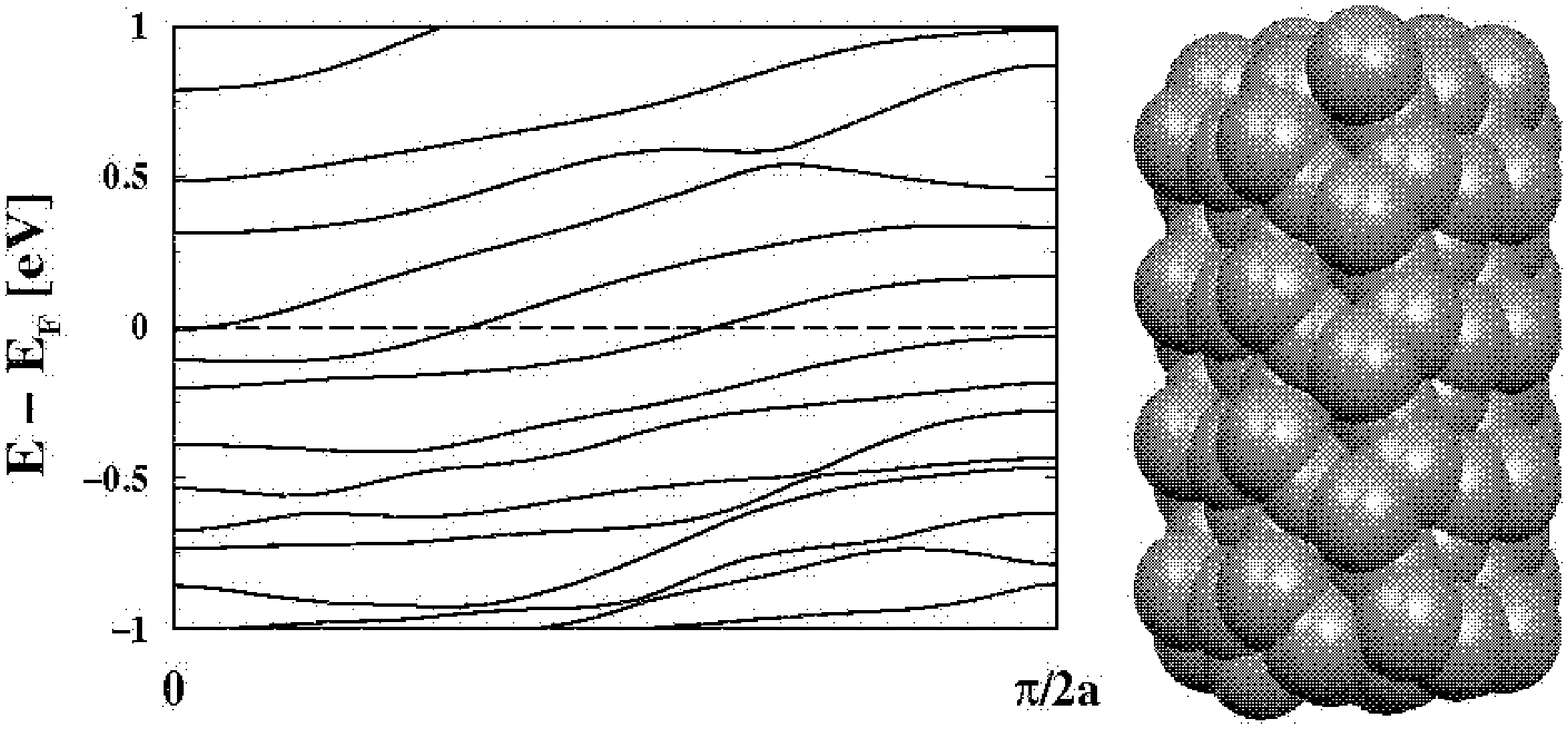}
\end{center}
\caption{(Color online) $\langle 100 \rangle$ SiNWs of $\sim11$~\AA\ 
         diameter: side view of the reconstructed (pseudo) \{100\} facet.}
\label{fig:100_11_relax_side}
\end{figure}

The most stable reconstruction that we have found for the $\sim11$~\AA\ 
wire is shown in Fig.~\ref{fig:100_8_relax_side}(a)~\cite{lda_gga}. 
It follows the same pattern of the previous wire, featuring one-single 
row of dimers. The differences between these two cases are better 
understood looking at Fig.~\ref{fig:100_thin_sec}(a-c), where the 
unrelaxed geometries are displayed. The thinnest wire can form one only 
row of dimers, involving all of the surface atoms of the \{100\} 
facet [panel (a)]. The $\sim 11$~\AA\ SiNW can also form one only 
row of dimers, but it leaves one unpaired row of Si atoms [panel (b)]. 
This has two notable effects: it results in a much more asymmetric in-plane 
relaxation [see Fig.~\ref{fig:100_thin_sec}(b)] and leaves a higher
density of dangling bonds. The effects are clearly visible also in the side
view of Fig.~\ref{fig:100_8_relax_side} and \ref{fig:100_11_relax_side}.

\begin{figure}[h]
\begin{center}
\epsfxsize=14cm
\epsffile{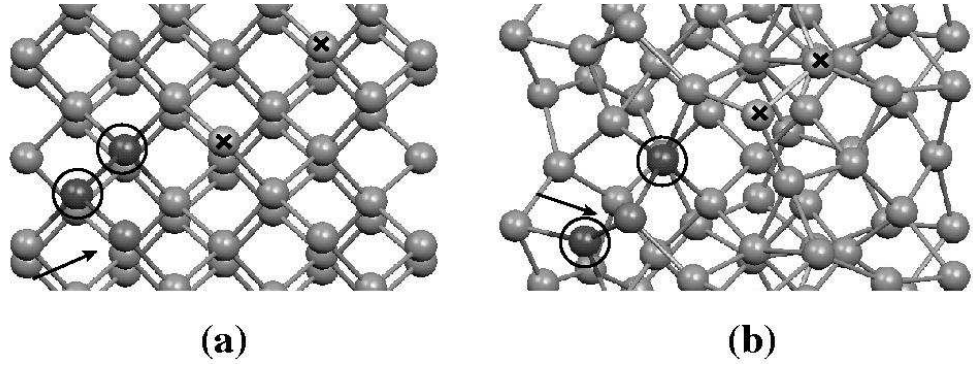}
\end{center}
\caption{(Color online) The mechanism that rules the facet relaxation 
         when not all the dangling bonds can form dimers. The {\em gray}
         atom (indicated with an arrow) is unpaired and cannot 
         form a dimer; to reduce the number of dangling bonds 
         breaks a bond of the underlying layer ({\em red} atoms 
         (indicated with a circle) and enters it. The {\em golden} 
         atoms (indicated with a cross) form a surface dimer according 
         to the conventional mechanism of Si(100)-like surfaces.}
\label{fig:100_11_relax}
\end{figure}

The asymmetry in the relaxation originates from the high reactivity of the
surface atoms of the unpaired row. These atoms cannot form dimers, but
their double dangling bonds would make the system unstable. The situation 
is depicted in Fig.~\ref{fig:100_11_relax}: the gray atom (indicated with 
an arrow) eliminate one dangling bond by breaking a Si-Si subsurface bond 
(red atoms, also indicated with a circle in the picture) and {\em entering} 
it. The golden atoms (indicated with a cross in the picture) therein are 
strictly equivalent to the gray atom, but they can combine and form a 
dimer, according to the conventional mechanism of Si(100) surfaces.



The next $\langle 100 \rangle$ SiNW that we analyze has already 
been described in detail elsewhere~\cite{noi_prl}. In this case the 
diameter ($\sim15$~\AA) starts to be sufficiently thick to permit 
to distinguish clearly between a bulk region, consisting of atoms
tetra-coordinated, and a surface region, consisting of atoms with
lower coordination: the surface region is the one where the 
reconstruction occurs, while in the bulk region the atoms maintain 
the ideal position of the Si lattice.

In Ref.~\onlinecite{noi_prl} we have shown that this wire sustains 
two different competing reconstructions, which only differ by 3~meV/atom. 
While one of them is strongly metallic, with four bands crossing the 
Fermi level, the other is at most semi-metallic.
In this case the assignment of a metallic character, at least to 
one of the two stable phases, is robust against failures of DFT,
because two of the four metallic states are degenerate and thus
cannot be shifted by a scissor operator without violating 
correct electron counting~\cite{noi_prl}.

\begin{figure}[h]
\begin{center}
\epsfxsize=14cm
\epsffile{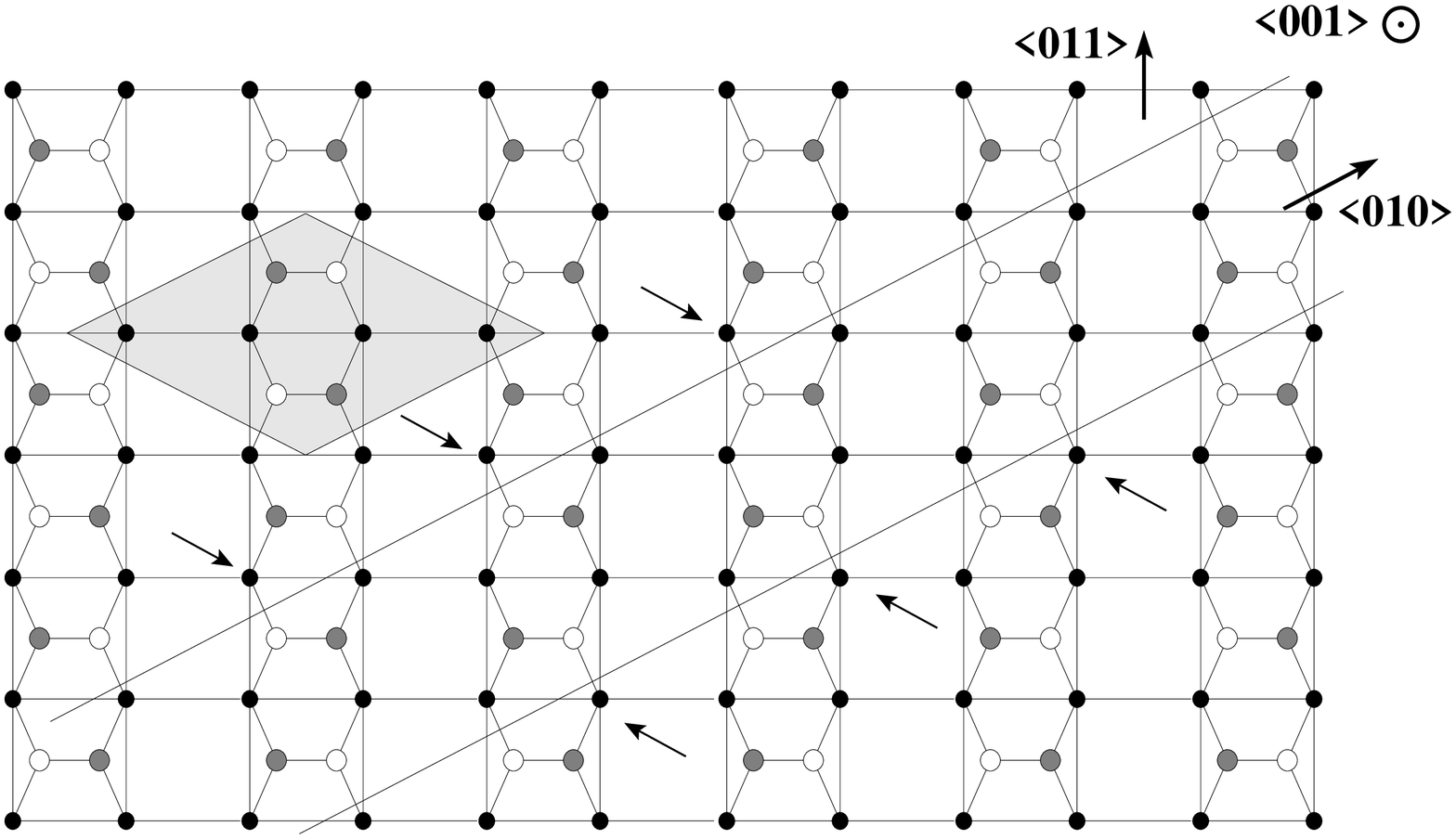}
\end{center}
\caption{Top view of the Si(100) $c(4\times2)$. White and gray balls 
         represent the (asymmetric) top dimers, dark balls are the 
         second layer atoms. A strip equivalent to the \{100\} facets 
         of the $\sim 15$~\AA\ $\langle 100 \rangle$ SiNW is drawn. 
         Each of the atoms represented by white balls have only one 
         second layer atom where to rest onto (the missing ones are 
         indicated with an arrow). This induces a tilt of the dimer
         axis when it relaxes, thus achieving a different packing 
         density than the infinite Si(100) surface.}
\label{fig:facet_surf}
\end{figure}

This opens up a fundamental question: how can surface states
of a \{100\} facet be metallic, while the analogous
reconstruction of the Si(100) surface is semiconducting?
In Fig.~\ref{fig:facet_surf} we show a top view of the $c(4\times2)$
reconstruction of the Si(100) surface; it can be seen that by
{\em cutting} a strip out of it, one recovers the metallic
reconstruction of the \{100\} facet (the one referred to as
{\em symmetric} in Ref.~\onlinecite{noi_prl}). However, the 
coordination of the boundary dimers change, because they now 
miss one of the {\em rest} atoms (some of them are indicated
with an arrow in Fig.~\ref{fig:facet_surf}).
This in turn alters the packing density of the surface dimers,
especially in the case of a nanometric size SiNW where {\em all}
dimers are at the facet boundary. As the diameter of the wire 
is increased, a growing number of dimer rows will recover the
coordination that they have on the Si(100) surface.
Hence, when the fraction of dimers at the facet boundary becomes
negligible a semiconducting behavior is expected.

\begin{figure}[h]
\begin{center}
\epsfxsize=14cm
\epsffile{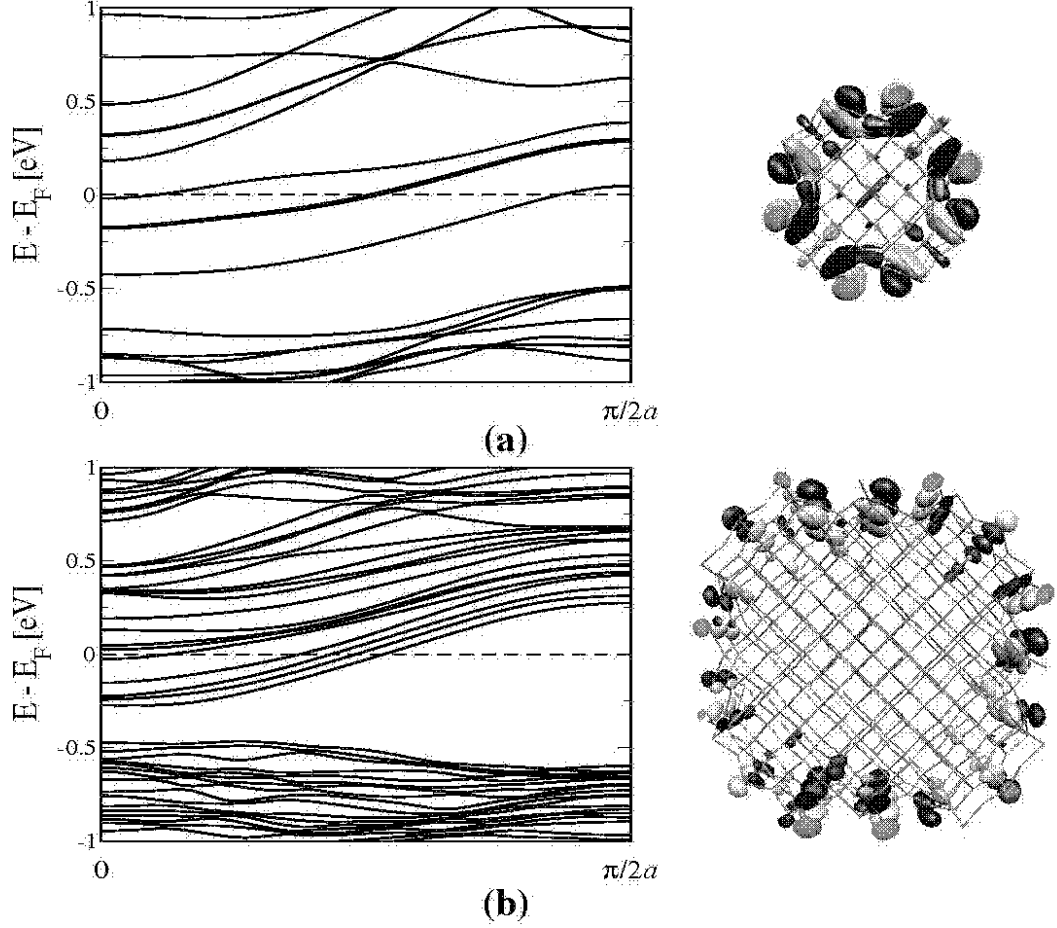}
\end{center}
\caption{(Color online) Band structure diagram of (a) the 15~\AA\ 
         SiNW of Ref.~\onlinecite{noi_prl} and (b) the 27~\AA\ SiNW.
         The comparison between the two panels shows the
         trend of band-gap opening as the diameter is increased.} 
\label{fig:100_15-27_bands}
\end{figure}

The thicker $\langle 100 \rangle$ SiNW that we have studied has a 
diameter of approximately 27~\AA. Up to this point we have distinguished
between wires that have an even or an odd number of rows of dangling
bonds in the \{100\} facets, because the different way of reducing
the number of dangling bonds, i.e. forming dimers or through the
mechanisms of Fig.~\ref{fig:100_11_relax}, proved to be the most 
prominent factor ruling the reconstruction. In the wires discussed 
so far the dimers formed on the facet were characterized by an 
altered packing density (the mechanism is illustrated schematically 
in Fig.~\ref{fig:facet_surf}), which ultimately results in 
the metalization. In this wire three row of dimers are formed on 
each \{100\} facet, so it is the first among the SiNWs  
that we have considered that features at least one row of dimers
which has the same nearest-neighbor coordination than the infinite
Si(100) surface. This situation is revealed by the band structure
of Fig.~\ref{fig:100_15-27_bands} where, though still being metallic, 
the band-gap starts opening, as can be seen by comparison with the 
15~\AA\ wire~\cite{noi_prl}. Another evidence of the size effect 
can be appreciated in the side panels of the same figure. There, we 
have plotted the wave function of the lower metallic state. It can 
be seen that the overlap between surface dangling bonds decreases 
for the larger diameter, indicating a higher localization of the
surface state. 

\begin{center}
$\sim$
\end{center} 

Summarizing, the metalization of $\langle 100 \rangle$ SiNWs is 
due to the partial distortion of the Si dimers on the \{100\}
facets which alters the overall packing density respect to the
infinite Si(100) surface. When the fraction of dimers with the
same coordination of the infinite surface increases, {\em i.e} 
when the Si(100) packing density is recovered, the SiNW
develops a semiconducting character.

\subsection{$\langle 110 \rangle$ SiNWs}
\label{sub:110}

\begin{figure}[h]
\begin{center}
\epsfxsize=14cm
\epsffile{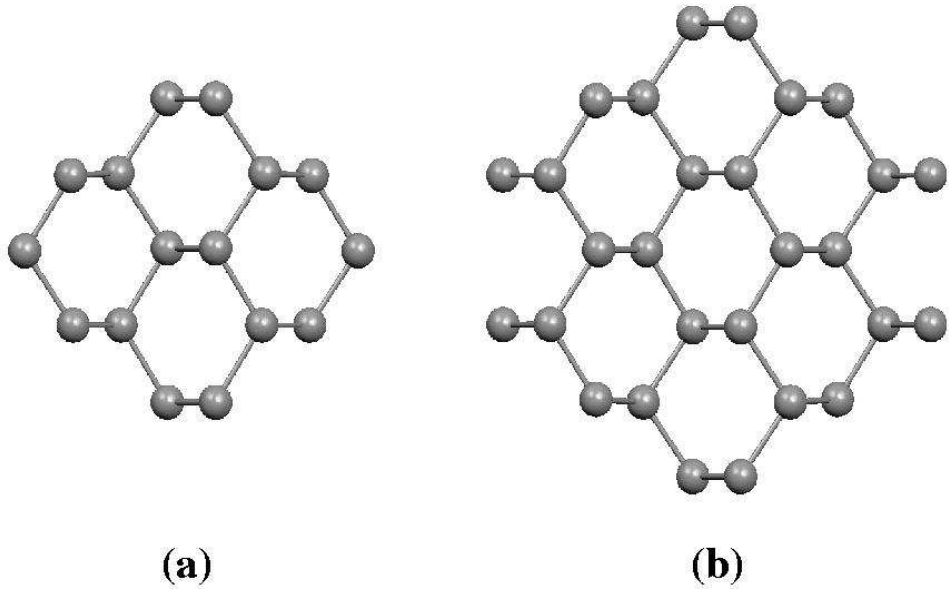}
\end{center}
\caption{(Color online) Possible geometries considered for 
         $\langle 110 \rangle$ SiNWs. The section that features 
         a central hexagon [shown in panel(b)] turned out to be 
         the most stable.}
\label{fig:110_top_geometries}
\end{figure}

\begin{figure}[h]
\begin{center}
\epsfxsize=14cm
\epsffile{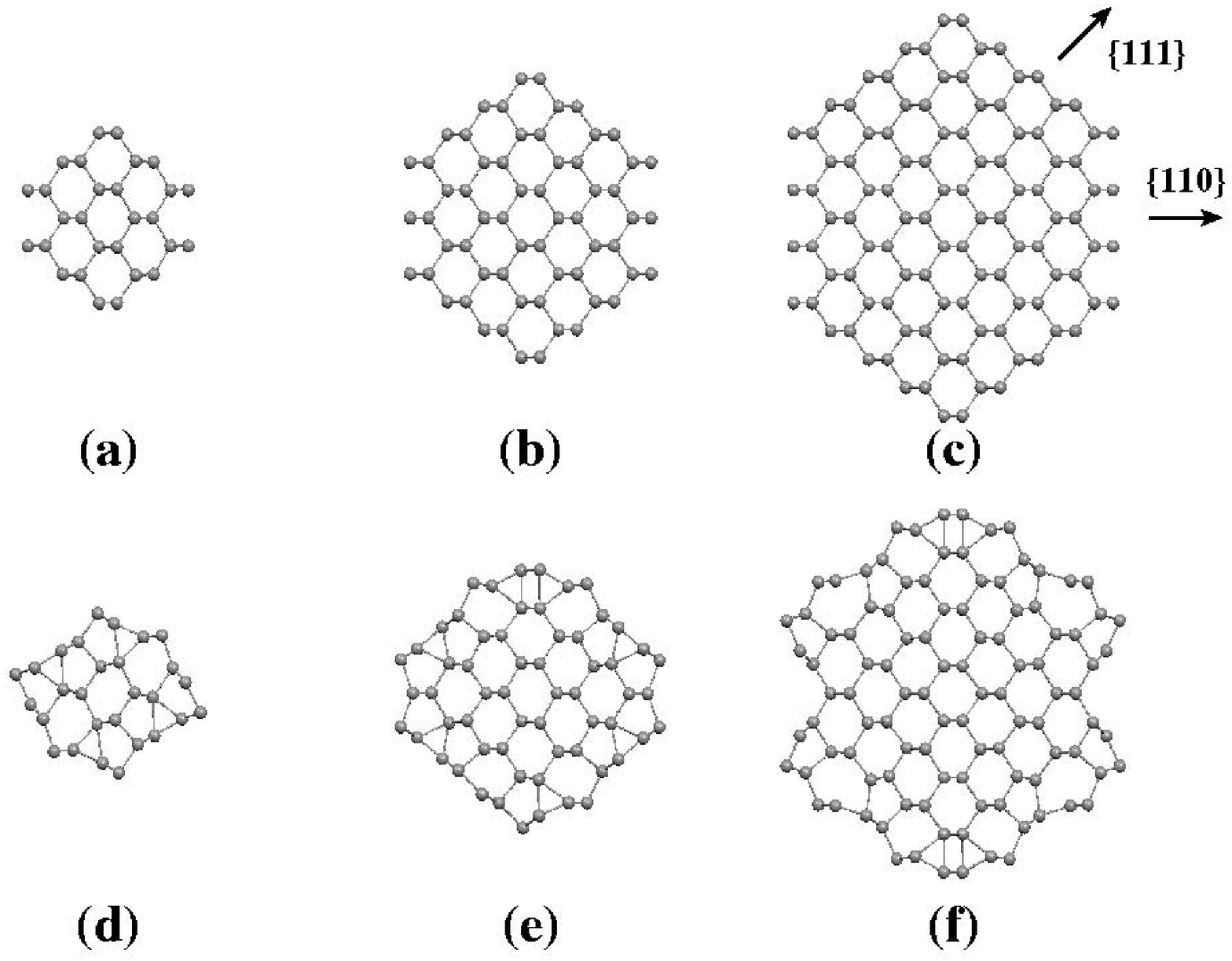}
\end{center}
\caption{(Color online) Section views of the unrelaxed [(a) to (c)] 
         and relaxed [(d) to (f)] $\langle 110 \rangle$ SiNWs with 
         a diameter of $\sim 12$\, $\sim 20$ and $\sim 26$~\AA}.
\label{fig:110_top_unrelax}
\end{figure}

In the case of $\langle 110 \rangle$ SiNWs the shape of the wire section 
is automatically defined by the hexagonal symmetry of the [110] plain of 
bulk Si. The hexagonal shape of the section of $\langle 110 \rangle$
SiNWs has been also confirmed unambiguously by scanning tunneling 
microscopy~\cite{ma} and transmission electron microscopy~\cite{wu} images.
It should be stressed that in this case, contrary to $\langle 100 \rangle
$ SiNWs, the angles between vicinal facets are intrinsically smooth. 
Hence, the only issue concerning the cross-section consists
in choosing between two possible configurations, shown in
Fig.~\ref{fig:110_top_geometries}: one with rings of hexagons
organized around a single hexagonal channel and another without
the central hexagon. The tests that we have performed
indicated that the arrangement in Fig.~\ref{fig:110_top_geometries}(b)
is favored. Therefore, this is the pattern that we have adopted to
generate SiNWs with growing diameters (Fig.~\ref{fig:110_top_unrelax}).
In these SiNWs the unrelaxed lateral surface is made up of \{111\}
and \{110\} facets. Therefore, the competition mechanism that rule
the reconstruction is expected to be markedly different to the case
of $\langle 100 \rangle$ wires.

\begin{figure}[h]
\begin{center}
\epsfxsize=14cm
\epsffile{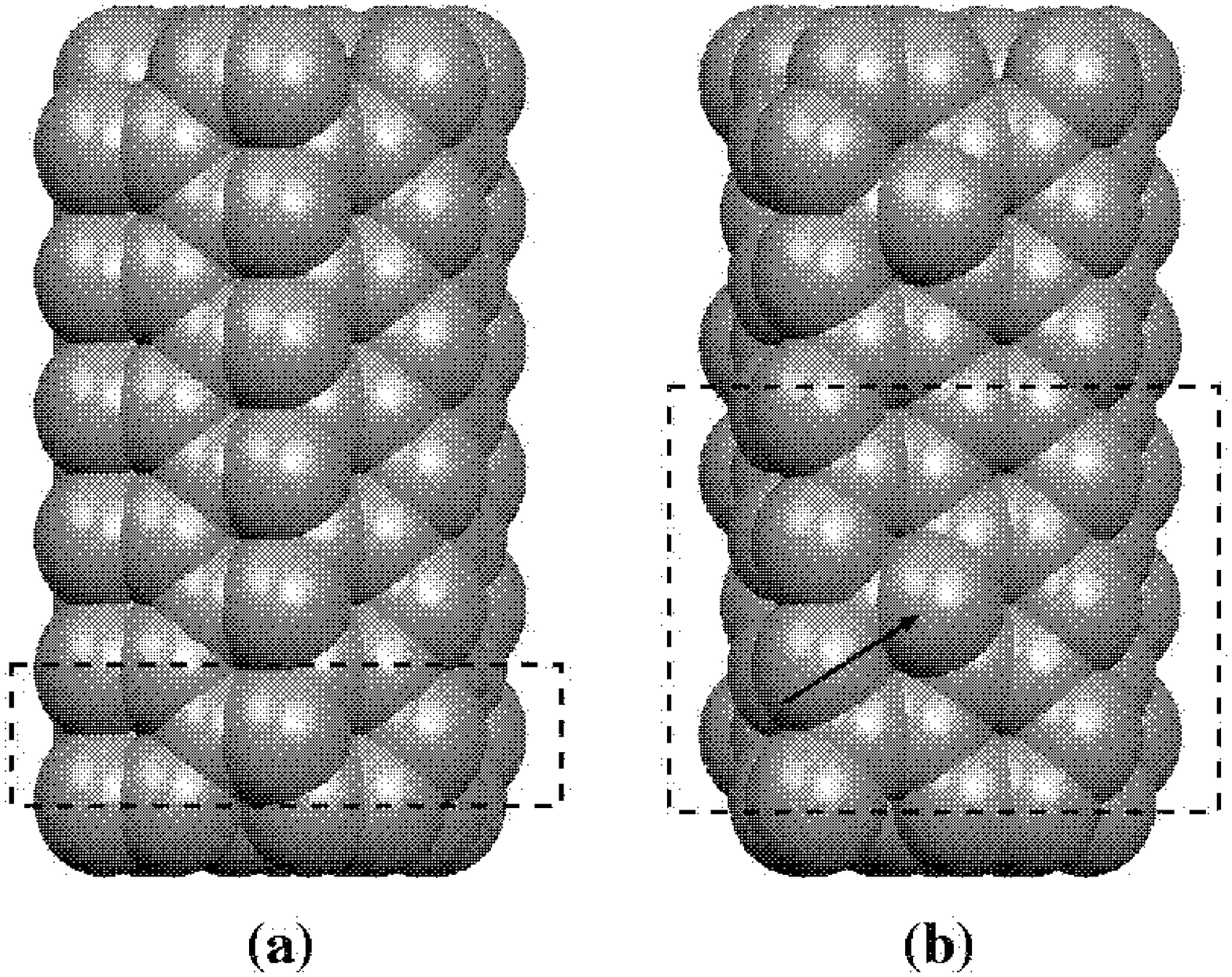}
\end{center}
\caption{(Color online) \{111\} facets of the two stable reconstructions 
         that we have 
         found for the 12~\AA\ $\langle 110 \rangle$ SiNW. The different 
         unit cells of the facet are indicated. The reconstructions have 
         extension of (a) $c$ and (b) $3c$, where $c$ is the axial 
         lattice parameter of the SiNW. In (a) the reconstruction features 
         one row of aligned atoms, while in (b) only one atom each three
         (indicated with an arrow) protrudes outward.}
\label{fig:110_12Ang_2phases_side}
\end{figure}

The thinnest $\langle 110 \rangle$ SiNW that we have studied turns
out to have two competing reconstructions, which are shown in 
Fig.~\ref{fig:110_12Ang_2phases_side}. After the relaxation it is hardly 
possible to still distinguish facets with a well-defined crystallographic 
orientation [see the top view of Fig.~\ref{fig:110_top_unrelax}(d)].
The difference between to the reconstructions found emerges at the
\{111\}-like facets, where the periodicity of the protruding atoms
change, as it can be observed in the space-fill representation 
of Fig.~\ref{fig:110_12Ang_2phases_side}. The difference in energy
is negligible, so that both reconstructions are expected to
occur with the same probability. Most important, both 
phases are semiconducting with a small band-gap, approximately
one third of its bulk value. 


The $\sim 20$~\AA\ has similar features. It presents two reconstructions
with a level of stability of the same order (the difference is around 
3~meV/atom). In this case too, the rows of protruding atoms can
arrange in different ways: one single row in the case of the
reconstruction with the smallest motif; a second row which alternates
protruding to non-protruding atoms in the other reconstruction.
Contrary to some of the cases described in Sec.~\ref{sub:100}, the
differences in the reconstruction alter only marginally the band
structure and a semiconducting character can be safely attributed 
to both phases.

\begin{figure}[h]
\begin{center}
\epsfxsize=14cm
\epsffile{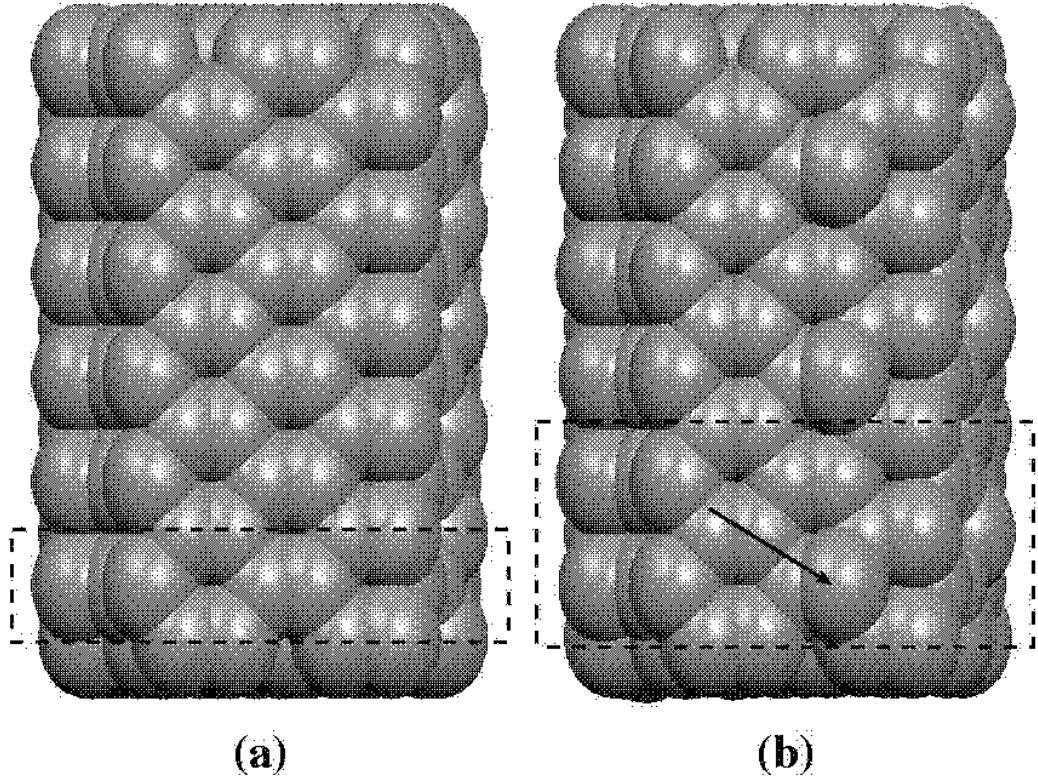}
\end{center}
\caption{(Color online) \{111\} facets of the two stable reconstructions 
         of the 20~\AA\
         $\langle 110 \rangle$ SiNW. The different unit cells of the 
         facet are indicated. The reconstructions have extension of 
         (a) $c$ and (b) $2c$, where $c$ is the axial lattice parameter 
         of the SiNW. In (a) the reconstruction features one row of 
         aligned atoms, while in (b) only one atom each two (indicated 
         with an arrow) protrudes outward.}
\label{fig:110_20Ang_2phases_side}
\end{figure}

Curiously, it turns out that the number of protruding 
atoms is always a constant. In the smallest reconstruction 
[Fig.~\ref{fig:110_20Ang_2phases_side}(a)] none of the 
atoms protrude outward, while all of them do in the 
opposite facet (not shown here). The opposite point of 
view was chosen for the unit-cell reconstruction of 
Fig.~\ref{fig:110_12Ang_2phases_side}(a), where all the
atoms are protruding outward, while none of them does 
in the opposite facet. In the same way, the more varied 
motif of the extended reconstructions of 
Fig.~\ref{fig:110_12Ang_2phases_side}(b) and 
Fig.~\ref{fig:110_20Ang_2phases_side}(b) is echoed in the 
opposite facet, where a depressed atom corresponds to a 
protruding atom and viceversa.


The largest $\langle 110 \rangle$ SiNW that we have analyzed has
a diameter of $\sim 26$~\AA. A significant difference with the
thinner wires described here, is that the diameter has reached a critical
size so that not only \{111\}, but also \{110\} facets are present.
The latter reconstructs forming a long trough all along the wire
axis, while the \{111\} have a patter similar to those seen so far,
with a row of protruding atoms.

A common feature that clearly emerges is that, although there is a 
dependence of the band-gap size on the wire diameter, in none of the
$\langle 110 \rangle$ SiNWs that we have studied the facet reconstruction 
gives rise to surface metallic states like in the case of wires grown 
along the $\langle 100 \rangle$ direction. From the discussion of the
results it appears that these wires are dominated by \{111\} facets
and \{110\} facets form only on thick wires.

The found geometries -~and their semiconducting character~-
agree with the $\pi$-bonded chain model typical of the
$2\times1$ reconstruction of the Si(111) surface~\cite{chadi111,
pandey,northrup,feenstra}. The unreconstructed surface (the facet
in the case of the wire) features one dangling bond per atom
and the system spontaneously reduces its symmetry to lower the
energy of the half-occupied states. This is possible because
the surface dangling bonds reside on nearest-neighbor atoms
and the bond-orbitals are close enough to interact
significantly stabilizing the surface and introducing a
band-gap between occupied and unoccupied states~\cite{pandey}.
Other reconstructions, notably
the $7\times7$, are known to be metallic due to the presence
of adatoms that have not been considered in the present study.
Adatoms could lead to metallic -~and possibly more stable~- phases
also in the case of nanowire facets and their influence on the electronic
and geometrical structure of the reconstruction should be addressed.
However, under a certain diameter the facets are not large 
enough to host such extended reconstructions, but the existence of 
adatom phases cannot be ruled out.

\begin{figure}[h]
\begin{center}
\epsfxsize=14cm
\epsffile{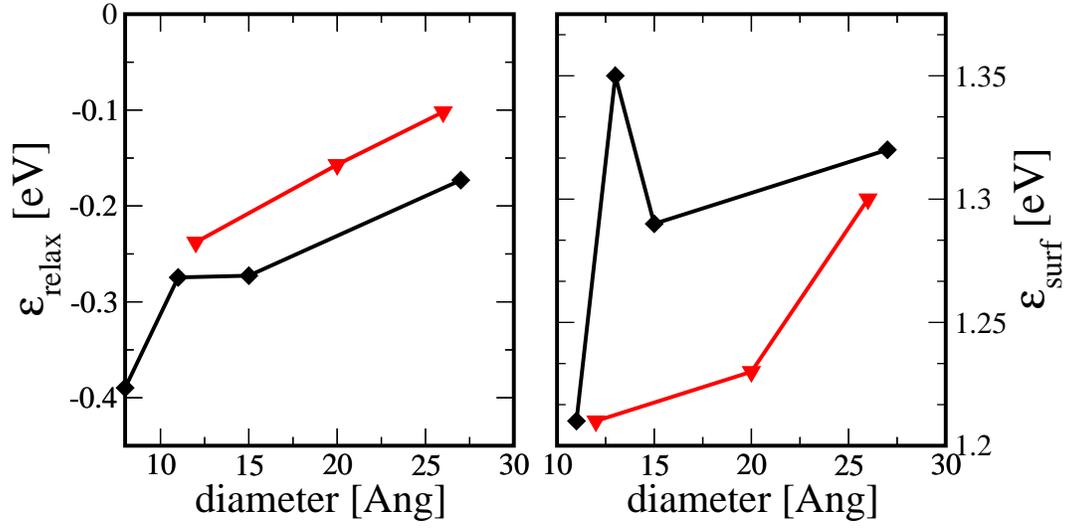}
\end{center}
\caption{(Color online) Energy gained with the relaxation, Eq.~\ref{eq:rel_en} 
         (left panel) and surface energy, as defined in 
         Eq.~\ref{eq:surf_en} (right panel).}
\label{fig:surf_en}
\end{figure}

Further insight on the mechanisms that rule the surface physics of 
these wires can be obtained from the analysis of Fig.~\ref{fig:surf_en}.
In the left panel we have plotted the energy gained with the relaxation:
\begin{equation}
\epsilon_\textrm{rel} = ( E^\textrm{rec}_\textrm{tot} - 
E^\textrm{unrec}_\textrm{tot} ) / n,
\label{eq:rel_en}
\end{equation}
where $E^\textrm{rec}_\textrm{tot}$ and $E^\textrm{unrec}_\textrm{tot}$ 
are the total energies of the reconstructed and unreconstructed geometry 
(shown in Figs.~\ref{fig:100_thin_sec} and \ref{fig:110_top_unrelax})
of a wire of $n$ atoms. The first thing to observe is 
that with the relaxation $\langle 110 \rangle$ wires lower their 
energy less than $\langle 100 \rangle$. This is in agreement with 
the reconstruction mechanisms previously discussed: $\langle 110 
\rangle$ wires simply rearrange the topology of the surface dangling 
bonds to allow the formation of a $\pi$-bonded chain; $\langle 100 
\rangle$ wires, on the other hand, go through a much stronger 
reconstruction of the facet, involving the formation of new bonds. 

Both curves are fairly linear, with the remarkable exception of
the 11~\AA\/ $\langle 100 \rangle$ SiNW. In the above discussion 
(see Sec.~\ref{sub:100}) we have suggested that facets of $\langle 
100 \rangle$ SiNWs with an even number of dangling bond rows should 
be favored. In the opposite case (odd number of dangling bond rows, 
like the 11~\AA\ wire), the system has to spend a larger amount of 
structural energy to readjust the facet geometry, because it cannot 
form ordered rows of dimers. This intuitive idea is confirmed by 
Fig.~\ref{fig:surf_en}(a) and becomes clearer by defining the surface 
energy as: 
\begin{equation}
\epsilon_\textrm{surf} = ( E_\textrm{tot} - n_\textrm{bulk} 
                         \epsilon_\textrm{bulk} ) / n_\textrm{surf},
\label{eq:surf_en}
\end{equation}
where $\epsilon_\textrm{bulk}$ is the energy per atom of bulk Si and 
$n_\textrm{bulk}$ and $n_\textrm{surf}$ the number of bulk and surface 
atoms, respectively~\cite{surf_en}. In this way, in the graph of 
Fig.~\ref{fig:surf_en}(b) we isolated pure surface contributions and 
the {\em anomaly} of the 11~\AA\ wire emerges in a more evident way.

As a concluding remark, we discuss the cohesion energies defined 
as~\cite{io_prb}:
\begin{equation}
\epsilon_\textrm{coh} = ( E_\textrm{tot} - n \epsilon_\textrm{Si} ) / n,
\label{eq:coh_en}
\end{equation}
where $E_\textrm{tot}$ is the total energy of the wire, $n$ is the 
number of atoms and $\epsilon_\textrm{Si}$ the energy of the isolated 
Si atom. The cohesion energy~\cite{io_prb}, being directly derived from 
the total energy, allows to compare the relative stabilities of the 
systems studied and which structures will be favored, in absence of 
other constraints, in the growth process.
The smoother angle between vicinal facets favors $\langle 110 \rangle$ 
growth orientation to the sharper angle of $\langle 100 \rangle$ SiNWs. 
When the diameter increases the difference tends to disappear: 
$\langle 100 \rangle$ wires reach a critical cross-section that 
allows to form smoother transition facets, while thick $\langle 110 
\rangle$ wires start to develop their own \{100\} facets, which are 
expected to contribute with a surface energy very similar to the 
\{100\} facets of the $\langle 110 \rangle$ SiNWs. 

\begin{center}
$\sim$
\end{center}

Summarizing, the $\langle 110 \rangle$ SiNWs that we have
analyzed turned out to be always semiconducting. 
The dangling bonds on the \{111\} facets cannot give rise to the 
formation of dimers because, due to the different symmetry
of the cleavage, they are second neighbors. Incidentally, if they
could form dimers this would result in a complete passivation
of the wire because each atom on the facet surface has only one
dangling bond per atom. However, the dangling bonds are close
enough to interact and to form a $\pi$-bonded chain typical
of Si(111) $2\times1$ surfaces.

\section{Conclusions}
We have performed first-principles electronic structure calculations
to study the surface reconstructions of $\langle 100 \rangle$ and 
$\langle 110 \rangle$ of different diameters.

SiNWs grown along the $\langle 110 \rangle$ axis are dominated by
\{100\} facets that exhibit the typical reconstruction pattern of
the Si(100) surface, consisting in the formation of rows of buckled 
dimers. In the case of the thinner wires, the altered packing density 
of the facet dimers gives rise to delocalized surface states that
result in a metalization of the surface. Increasing the diameter,
the coordination of the infinite Si(100) is recovered. The largest
$\langle 100 \rangle$ SiNW studied, though still metallic,
gives a clear indication of a band-gap opening.  

The case of SiNWs grown along the $\langle 110 \rangle$ axis is
different. Thanks to a different cross-section shape, the 
neighboring facets can match through a smooth angle and they are not 
significantly distorted with respect to their homologous infinite
surfaces. Hence, the \{111\} facets that dominate the
reconstruction follow the $\pi$-bonded chain model of Si(111) 
surfaces, conferring to the wire a semiconducting character.
The band-gaps are found to be smaller than the bulk value
as an effect of the formation of the $\pi$-surface states, while
as it is well-known, H-passivated wire of similar sizes result
in widened gaps as a consequence of quantum confinement.
Analysis of the energetics of the wires indicates that at small
diameters the growth of $\langle 110 \rangle$ SiNWs is favored
over $\langle 100 \rangle$ SiNWs.
\label{sec:conclusions}

\begin{acknowledgments}
R.R. acknowledges the financial support of the Generalitat de Catalunya
through a {\sc Nanotec} grant and of the Juan de la Cierva programme
of the Spanish Ministerio de Educaci\'{o}n y Ciencia.
Computational resources at the Centre Informatique National de 
l'Enseignement Sup\'erieur and the Centre de Calcul Midi-Pyr\'en\'ees 
are gratefully acknowledged.
\end{acknowledgments}

\newpage

\begin{table}
\begin{center}
\begin{tabular}{lcc}
\hline\hline
{\bf Orientation} & {\bf diameter [\AA]} & {\bf $E_\textrm{coh}$ [eV]}\\
\hline
{\bf $\langle 100 \rangle$} &    & \\
                            &  8 & -3.75 \\
                            & 11 & -3.86 \\
                            & 15 & -3.99 \\
                            & 27 & -4.16 \\
\hline
{\bf $\langle 110 \rangle$} &    & \\
                            & 12 & -3.93 \\
                            & 20 & -4.08 \\
                            & 26 & -4.14 \\
\hline\hline
\end{tabular}
\end{center}
\caption{Cohesion energies of the SiNWs studied. For thin diameters
         $\langle 110 \rangle$ are favored, but as the size is
        increased the difference tends to disappear.}
\label{tab:coh}
\end{table}

\clearpage

\clearpage

\clearpage

\clearpage

\clearpage

\clearpage

\clearpage

\clearpage

\clearpage

\clearpage

\clearpage

\begin{thebibliography}{99}
\bibitem{lda_gga} For this wire we have also found a different 
                  reconstruction. Though stable, it turned out 
                  to be more energetic than the one presented
                  and hence, it is not discussed here.
\end{thebibliography}
\end{document}